\documentstyle[prl,aps]{revtex}
\begin{document}

\title{On the Nature of the Compact Objects in the  AGNs}

\author{Abhas Mitra}

\address{
Theoretical Physics Division, Bhabha Atomic Research Center,\\
Mumbai-400085, India\\ E-mail: amitra@apsara.barc.ernet.in}


\maketitle


\begin{abstract}
If the cores of Galaxies and AGNs comprise Supermassive Stars (SMSs)
(Hoyle and Fowler 1963),  nuclear energy generation may not be ignited for
masses $M > 6\times 10^4 M_\odot$, and highly supermassive stars are
likely to generate their luminosity, $L_{KH}$ slow gravitational
contraction. We show that for such massive Newtonian SMSs, both  the value
of $L_{KH}$ and the the accretion luminosity, $L_{acc}$, could be much
less than the appropriate Eddington value. Such highly SMSs will have a
surface density as low as $10^{-14}$ g/cm$^3$ and they will not behave
like a compact object with a ``hard surface'. For all such reasons, it is
indeed possible to have $L_{SMS} \ll L_{ed}$. And this effect may be
incorrectly interpreted as the evidence for the existence of central Super
Massive Black Holes with Event Horizons.

In fact, in a very detailed work, and from various angles, we have
recently shown that, the General Theory of Relativity does not permit the
occurrenes of ``trapped surfaces'' and Black Holes (Mitra 1998a,
gr-qc/9810038).
\end{abstract}


\section{Introduction}
One of the cornerstones of modern physics and astronomy is the idea of
the Black Holes (BH) in which everything can enter but from which nothing can come
out, atleast at the classical level. Black holes are supposed to exist as
the compact object in many X-ray binaries, in the center of core collapsed
star clusters and in the core of the socalled
Active Galactic Nuclei (AGN) as well as in the core of many normal
galaxies, like the Milky way. They may exist in isolation too, but it is
difficult to detect such isolated BHs. In this work, we will focus
attention on the supposed Galactic core BHs. Despite such popular notions,
 it is simultaneously acknowledged that most of the evidences for the
existence of BHs are highly circumstantial, and, what one observes is
actually  Massive Compact Dark Objects (MCDO). For instance, it is widely
believed that the center of our galaxy harbors a BH of mass $M \sim 10^6 M_\odot$
having a radius $R_{BH} \sim 3\times 10^{11}$ cm. But actually the spatial
resolution with which we can scan this region is $\sim 0.1~pc \sim 3\times
10^{17} ~cm \sim 10^{6} R_{BH}$, and thus, as such, we can rule out the
possibility that the core may actually comprise densely packed X-ray
binaries, Wolf-Rayet stars or other massive stellar cores. In the course
of time such objects might also form a single self-gravitating entity
called Super Massive Star (SMS, Hoyle and Fowler 1963). The same is true
for the AGNs too although the intraday variabilites observed in many AGNs
would yield a tighter limit on the core size $R < 10^{15}$ cm (however,
note that the estimates of core masses are in general uncertain atleast by
a factor of $\sim 10^2$).

Even if such cores are accepted to be massive BHs, they are likely to have
been formed by a preceding phase as SMSs (Begelman and Rees 1978). Thus at a
given epoch, many such cores must be SMSs. It is believed that the
unassaiable evidence in favor of BHs and their Event Horizons arise from
the fact that accretion luminosity $(L_{acc})$  from many AGNs or
galactic cores is insignifican compared to the corresponding Eddington
values ($L_{acc} \ll L_{ed}$). This could be so because for weak or
excessively high accretion rates, the flow could be advection dominated
with small radiative efficiency. As a result, for spherical accretion, not
mediated by a disk, most of the accreted energy and mass will be lost
inside the event horizon. However, without (first) entering into a debate
on the possible existence BHs, we will show below that for SMSs too it is
possible to have $L_{acc} \ll L_{ed}$. Thus {\em  Super Massive Stars too may
emulate the Mass-Energy gobbling properties of a BH}.

\section{Supermassive Stars}
A SMS is  supported  almost entirely by its radiation pressure
\begin{equation}
p_r = {1\over 3} a T^4
\end{equation}
where $a$ is the radiation constant. On the other hand, by assuming the
plasma to be made of hydrogen only, the matter pressure is given by
\begin{equation}
p_m = n k T
\end{equation}
where $k$ is the Boltzman constant and n is the proton number density. The
structure of a Newtonian SMS is closely given by a polytrope of index 3,
and the ratio of matter pressure to gas pressure works out to be (Weinberg
1972):

\begin{equation}
\beta ={p_m\over p_g} \approx 8.3 \left({M\over M_{\odot}}\right)^{-1/2} =
8.3 \times 10^{-4} M_8^{-1/2}
\end{equation}

where $M = M_8 10^8 M_\odot$. This shows that a SMS should have a minimum
mass $\sim 7200 M_\odot$ (Weinberg 1972) in order to have a value of
$\beta <0.1$. A Newtonian SMS may be defined
as one for which the rest mass energy density of the plasma dominates over
the radiation energy density even though $p_m \ll p_r$, i.e., $ m_p n c^2
\gg 3 p_r$. It then follows that, the ``compactness'' of a Newtonian SMS
is very small (Weinberg 1972):
\begin{equation}
{2 G M\over R c^2} \ll 0.78
\end{equation}
In other words the surface red-shift of a Newtonian SMS is very small

\begin{equation}
z = \left(1 - {2GM\over R c^2}\right)^{1/2} -1 \ll 0.39
\end{equation}

In this limit of small $z$, we can approximate $z \approx GM/R c^2$.
Most of the stable (Newtonian) SMSs are likely to have $10^{-4} < z < 10^{-2}$ (Shapiro
\& Teukolsky 1983).
However, it is also possible to have real compact {\em Relativistic SMSs}
with high values of $z \sim 1$ for which {\em radiation dominates even in the
energy budget}. Such Relativistic  SMSs may be described by a relativistic
polytrope of degree 3 (Tooper 1964). But, in this work we shall discuss the case of Newtonian
SMSs only.

If a SMS is in {\em hydrosttic equilibrium}, its luminosity is close to
the corresponding Eddington value :
\begin{equation}
L = (1-\beta) {4 \pi c G M\over \kappa} \approx 1.26 \times 10^{46} M_8 ~erg/s
\end{equation}
where $\kappa $ is the Thompson opacity. And as is well known, this fact
was one of the basic reasons behind hypothesizing that the quasars could
be powered by SMSs (Hoyle and Fowler 1963). Note, the hydrostatic
equilibrium must be effected by the release of energy by nuclear fusion at
the center. But the efficiency for energy generation by hydrogen fusion is
only $\sim 0.7\%$, and given Eddington limited accrretion rate, the fusion
process can not deliver the necessary luminosioty for massive SMSs. In
fact, it is difficult to conceive of nuclear-fuel supported SMS for $ M >
6\times 10^4 M_\odot$ (Shapiro \& Teukolsky 1983). Thus, {\em more massive SMSs
are not in hydrostatic equilibrium and must be undergoing slow
gravitational contraction}.

\section{ Highly Supermassive Stars (HSMS)}
Now we shall estimate the luminosity of Newtonian HSMS releasing energy by
the Kelvin-Helmohltoz process and for which nuclear energy generation is insignificant:
First note that, the effective adiabatic index of a pure H-SMS is given by
(Weinberg 1972)
\begin{equation}
\gamma \approx {4\over 3} + {\beta \over 81}
\end{equation}
Then the Virial Theorem looks like

\begin{equation}
E_{in} + 3 (\gamma -1) E_g =0; \qquad or, E_{in} + (1 +\beta/27) E_g =0
\end{equation}

where $E_{in}$ is the total internal energy and $E_g$ is the
self-gravitational energy. On the other hand, the Newtonian total energy
of the star (polytrope of index 3) is

\begin{equation}
E_N = E_{in} + E_g = -{\beta \over 27} \mid E_g\mid = {-\beta \over 18} {G M^2 \over R}
\end{equation}

The K-H contraction luminosity is then given by

\begin{equation}
L_{KH} = -{dE_N\over dt} \approx {\beta \over 18} {GM^2\over R^2} v
\end{equation}

where $v \ll c$ is the rate of slow contraction. In the limit of small
$z$, by using Eqs. (3) and (5), the above expression can be rewritten in a physically significant manner:

\begin{equation}
L_{KH} = {\beta\over 18} {z^2 c^4 v\over G} \approx 8.3 \times 10^{-4}
{z^2 c^4\over 18 G} M_8^{-1/2}
\sim 6.25\times 10^{38} z_3^2 v M_8^{-1/2}~ erg/s
\end{equation}

where $z= z_3 ~ 10^{-3}$.
By comparing this pure gravitational contraction luminosity with the
corresponding Eddington value, we find that

\begin{equation}
{L_{KH} \over L_{ed}} \sim 5\times 10^{-8}~ z_3^2~ v~ M_8^{-3/2}
\end{equation}

Here, we emphasize that, the fact that {\em the system is out of hydrodynamical
equilibrium need not mean the system is undergoing free fall}, and the
value of $v$ could be much smaller than the corresponding free fall speed
$v_{ff} \approx \sqrt{2z} c$. The KH contraction may continue for thousand
of years and the value of $v$ could be as low as $few~ km/s$ for the initial phase.
Then we have

\begin{equation}
{L_{KH} \over L_{ed}} \sim 5\times 10^{-3}~ z_3^2~ v_1~ M_8^{-3/2}
\end{equation}

where $v= v_1 ~ 1 Km/s$.
Thus the {\em intrinsic KH luminosity of a HSMS could be insignificant compared
to the expected Eddington luminosity}.

\section{Accretion Luminosity}
We have recently shown that the GTR expression for accretion luminosity
from a ``hard surface'' (Mitra 1998b) is

\begin{equation}
L_{acc} = {z\over 1+z} {\dot M} c^2 \approx z {\dot M} c^2 \ll L_{ed}; \qquad if,~ z\ll
1
\end{equation}

In the limit of small $z$ the above general formula obviously yield the
Newtonian formula : $L_{acc} = {G M {\dot M}\over R}$. In this case, the
accretion efficiency $z$ could be much smaller compared to the
corresponding value of disk accretion onto a Schwarzschild BH, $\sim 5.7
\%$. Further, the actual accretion efficiency could be much smaller than
what is indicated by Eq. because a {\em HSMS does not really have a ``hard surface''}.

\section{Hard Surface ?}
The mean baryonic density of a SMS is

\begin{equation}
{\bar \rho} = {M\over (4\pi/3) R^3}= {3 z^3 c^6 \over 4\pi G^3 M^2} \sim
10^{-13} ~z_3^3 ~M_8^{-2} ~ g/cm^3
\end{equation}

And for a polytrope of index 3, the density of the external layers is
atleast one order smaller:

\begin{equation}
{\bar \rho}_{ex} \sim
10^{-14} ~z_3^3 ~M_8^{-2} ~ g/cm^3
\end{equation}

With the low luminosity inferred above, it may be found that, the
temperatures of these layers could be $< 1 eV$, and the gas is actually
partially ionized (heavy elements would be almost completely neutral) !

Then an incident electron  of energy $\sim 0.5 z MeV$ or a proton of
energy $\sim z GeV$ would primarily
undergo low energy ionization/nuclear losses. Even otherwise, at such low
densities, the test particle may penetrate deep inside the HSMS and any
photon that it might emit may be trapped in the general soup of photon and
plasma. Thus it is possible that $L_{acc} \ll L_{ed}$.

 And {\em this may be
mistaken as an evidence in favor of the existence of an Event Horizon}!!

\section{Discussion}
In this preliminary work we have outlined that Highly Supermassive stars
with low values of compactness, $z$, are likely to
undergo slow Kelvin-Helmoltz contaction. The resultant luminosity arising
from either this contraction process or accretion of surrounding gas
clouds, are likely to yield a
luminosity which is insignificant compared to the corresponding Eddington
value. And given such an observational result, in the absence of a serious
consideration of the physics of the HSMSs, one might be tempted to
describe the result as ``Experimental Discovery of Black Holes''.

On the other hand, in the relativistic regime (not described here), the
Supermassive stars may be very compact $z <0.615$ and have central
temperatures well above $T_c > 10^9$ K. The gravitational contraction
luminosity of
such stars may be released in the form of neutrinos. Such compact stars
will possess a ``hard surface'' and their accretion luminosity could be
comparable  to the corresponding Eddington
value.

In the
context of suspected stellar mass BHs, it may be reminded that it might be
possible to have novel kind of hadronic stars, called, Q-stars whose value
could be as high as $> 100 M_\odot$ (Bachall et al. 1989, Miller et al. 1998).
 Further, {\bf Q-stars may be formed at
sub-nuclear densities} and depending on the uncertain model QCD parameters,
have a wide range  of $1 < z \ll 1  $. In case, $z \ll 1$, one would have
$L_{acc} \ll L_{Ed}$, and again this may be mistaken as an evidence
for an ``event horizon''.

Finally, in a very detailed work (Mitra 1998b, gr-qc/9810038), we have
shown from all possible angles that BHs neither form in gravitational
collapse of baryonic matter, nor can  they be assumed to exist, in general.
The only exact solution of Einstein eq. which (apparently) shows the
production of BHs in spherical collapse is due to Oppenheimer and Snyder
(1939, OS). As is well known, they (apparently) showed that
 the collapse of a {\em homogeneous dust} of mass $M$ gives rise to a BH
of same mass in a proper time $\tau \propto M^{-1/2}$. However, what they
{\bf overlooked} in this historical paper is that the Schwarzschild time $t$ is
related by a relation of the kind (see Eq. 36 of OS):

\begin{equation}
t \sim \ln{y^{1/2} +1\over y^{1/2} -1} ; \qquad y \sim Rc^2/2GM
\end{equation}

In order that $t$ is a real quantity in the above expression, one must
have

\begin{equation}
y^{1/2} >1; \qquad or, ~~ {2GM\over Rc^2} <1
\end{equation}

 Then in order to reach the central
singularity, one must have

\begin{equation}
M\rightarrow 0; \qquad ~ as ~ R\rightarrow 0; \qquad ~ so ~that~ \tau \propto M^{-1/2} \rightarrow \infty
\end{equation}

 This means that the {\em BH is never produced and even if
it would be produced its mass must be zero}! We have then proved this
result in a most general fashion for an inhomogeneous dust as well as for a {\bf perfect fluid having arbitrary EOS
and radiative property} (Mitra 1998a, gr-qc/9810038).

Thus we may say that whether the cores of galaxies are SMSs or not, they
are certainly not BHs.


\end{document}